\documentclass[pre,reprint,floatfix,groupedaddress,superscriptaddress]{revtex4-1}
\pdfoutput=1
\usepackage[utf8]{inputenc}
\usepackage{amsmath, amssymb,mathtools,hyperref,subcaption}

\begin{document}\title{Arnol'd Tongues in Oscillator Systems with Nonuniform Spatial Driving}
\author{Alexander Golden}
\affiliation{Department of Physics, Boston University, Boston, Massachusetts 02215, USA}
\affiliation{Biological Design Center, Boston University, Boston, Massachusetts 02215, USA}
\author{Allyson E. Sgro}
\affiliation{Biological Design Center, Boston University, Boston, Massachusetts 02215, USA}
\affiliation{Department of Biomedical Engineering, Boston University, Boston, Massachusetts 02215, USA}
\author{Pankaj Mehta}
\affiliation{Department of Physics, Boston University, Boston, Massachusetts 02215, USA}
\affiliation{Biological Design Center, Boston University, Boston, Massachusetts 02215, USA}

\date{December 2020}
\begin{abstract}

Nonlinear oscillator systems are ubiquitous in biology and physics, and their control is a practical problem in many experimental systems. Here we study this problem in the context of the two models of spatially-coupled oscillators: the complex Ginzburg-Landau equation (CGLE) and a generalization of the CGLE in which oscillators are coupled through an external medium (emCGLE). We focus on external control drives that vary in both space and time. We find that the spatial distribution of the drive signal controls the frequency ranges over which oscillators synchronize to the drive and that boundary conditions strongly influence synchronization to external drives for the CGLE. Our calculations also show that the emCGLE has a low density regime in which a broad range of frequencies can be synchronized for low drive amplitudes. We study the bifurcation structure of these models and find that they are very similar to results for the driven Kuramoto model, a system with no spatial structure. We conclude by discussing the implications of our results for controlling coupled oscillator systems such as the social amoebae \emph{Dictyostelium} and populations of BZ catalytic particles using spatially structured external drives.
\end{abstract}
\maketitle

\section{Introduction}

Collective oscillations are ubiquitous in biology, playing important roles in a broad range of biological processes. Some notable examples include voltage oscillations in neural systems \cite{koizumi2010synchronization}, oscillatory insulin secretion in the pancreas \cite{benninger2008gap}, entrainment of circadian rhythms in cyanobacteria \cite{ito2007autonomous, van2007allosteric}, glycolytic oscillations in yeast \cite{dano1999sustained}, and internal and external biochemical oscillations during aggregation in social amoebae such as \emph{Dictyostelium discoidium}  \cite{gregor2010onset}. Across all of these systems, these oscillations play an important biological role in coordinating the behavior of cells across a population with each other as well as with external environmental stimuli. 

An important class of these biological oscillations is found in dynamical quorum sensing where the activity of individual cells is controlled by collective population-level oscillations \cite{de2007dynamical}. For example, in  \emph{Dictyostelium}, oscillations play an important role in mediating the collective response to changes in the environment by coordinating aggregation and then collective migration. The way that environmental perturbations or other drives affect the collective dynamics is an important element of these processes, not only in terms of understanding biological function but also for developing mechanisms to control experimental systems. Here we address an important aspect of control of collectively oscillating systems: the effectiveness of control stimuli on spatially structured cellular populations.

In the context of nonlinear oscillators, the question of control can be understood as the problem of synchronizing an oscillator to an external drive. Studies of how driving forces can control and coordinate oscillations date back to Huygens \cite{huygens1673christiani}, and have found applications in electronic systems, laser control, and many biological systems \cite{aranson2002world}. A common way of understanding synchronization to external drives is using ``Arnol'd tongues" which characterize the relationship between the amplitude of the drive $B$ necessary to induce synchronization and the difference between drive frequency and the oscillator's free running frequency, $\Delta\omega$. This relationship generally takes the form $B^2 = (\Delta\omega/W)^2$ when $B$ and $\Delta\omega$ are small, with $W$ the width of the Arnol'd tongue. This width $W$ can be interpreted as a representation of how easy the medium is to control since a larger $W$ means that a weaker drive can synchronize the oscillator to the same band of frequencies.

For a single oscillator, synchronization to an external drive is governed by two competing processes with different timescales. When the drive is weak the natural frequency controls the dynamics, and when the drive is strong enough the system synchronizes to the drive frequency. In more realistic oscillator models such as dynamical quorum sensing that have multiple timescales (for example, due to different media being produced or degraded at different rates), each of these timescales contribute to the effective ``natural frequency" of the oscillating medium and hence can play a role in determining which frequencies can be synchronized. We show below how introducing an external medium with its own dynamics introduces such additional timescales and how they affect synchronization, introducing new complexities to the control problem. Specifically, we focus on how coupled oscillators organized in space respond to spatially-structured external drives, taking inspiration from biological oscillations such as \emph{Dictyostelium}, as well as within and between microbial biofilms \cite{liu2015metabolic,liu2017coupling}. In \emph{Dictyostelium}, for example, the spatial organization of their self-driven oscillations plays a key role because these oscillations mediate the spatial partitioning of the population into multiple multicellular units which take on their own dynamics. 

The introduction of spatial organization in these coupled oscillator systems brings with it several complicating factors, such as the impacts of boundary conditions and the possibility of linear instabilities that can lead to spatial patterning or chaos \cite{cross1993pattern}. Boundary conditions in particular can play a subtle role in determining the types of solutions that are observed. Periodic boundary conditions, which are often practical for analytic calculations and simulations, generally cannot be realized experimentally. Even so, they can often provide useful intuition for the behavior of very large systems far away from the boundaries. Zero-flux boundary conditions are often the most appropriate choice for modeling biological oscillations. We demonstrate below that the choice of boundary conditions can play an important role in determining the persistence and stability of synchronization when populations are confined to live in a one-dimensional space.

In many systems of interest, such as cellular populations, nonlinear oscillators are not directly coupled but instead communicate with each other through an external medium that possesses its own dynamics. In the context of dynamical quorum sensing, the external media is often a single chemical species which cannot oscillate autonomously. Instead, the concentration of the external medium changes due to production of the chemical species by the oscillators and can also be degraded at some rate. This introduced three additional timescales to the dynamical quorum sensing problem: the difference in the oscillation period of the oscillators and the media,  a ``coupling" time which represents the time it takes for the medium to respond to the oscillators, and  a ``relaxation" time which sets the memory lifetime for the medium. The first timescale is a relative timescale between the oscillators and the external medium, and the latter two are characteristic of the external medium itself. The relationships between these timescales have been shown to play an important role in controlling the dynamics for mean field models of dynamical quorum sensing lacking spatial structure, and we show below how the relationships between these timescales describe the behavior of the system of oscillators.

One notable complication that can arise due to having multiple independent timescales in coupled oscillator systems is that the oscillation itself can become inhibited. This is known as ``amplitude death" in the literature on coupled oscillator systems \cite{strogatz1998death}. Amplitude death occurs when a time delay disrupts coupling between the oscillators to such a degree that the individual oscillators cease to oscillate and decay to zero amplitude. Previous work \cite{schwab2012dynamical,schwab2012kuramoto, noorbakhsh2015modeling} has studied the boundary of the amplitude death phase in oscillator systems coupled by an external medium in the mean field limit, and we will extend this to the case of spatial oscillators with a drive.


Motivated by the considerations discussed above, we investigate the use of spatially-structured external drives to synchronize and control cellular populations. We focus on two generic models that have been argued to be generic descriptions of oscillatory, diffusively coupled media near the onset of oscillations: the Complex Ginzburg-Landau equation (CGLE) and a generalization of CGLE to the case in which the oscillatory medium is coupled through an external medium (emCGLE) \cite{aranson2002world, tanaka2003complex, kuramoto1997phase}.  In section \ref{sec:CGL} we derive the Arnol'd tongue conditions describing the drive amplitude and frequency ranges for which the medium can be synchronized to the drive. Notably, we show that the spatial distribution of the drive controls temporal aspects of the synchronization process, in particular the wavelength of the drive can tune the frequency ranges to which the medium can synchronize. This section generalizes the results of Ref.  \cite{chate1999forcing} which focused on uniform drives for the CGL with periodic boundary conditions, and show that the range of frequencies which can be synchronized depends on the drive wavelength. In section \ref{sec:CGL0flux} we show that zero-flux boundary conditions can disrupt synchronization and induce higher-order spatial locking. In section \ref{sec:emCGL} we derive the Arnol'd tongue condition for the emCGLE and show that the wave number of the drive competes with memory effects in the medium itself for influence on the width and location of the Arnol'd tongue. We show that the emCGLE exhibits a low-density parameter regime, not found for the standard CGLE, in which the width of the Arnol'd tongue becomes very large and very low drive amplitudes can synchronize the medium to a wide range of drive frequencies.  We show that this new regime is related to the amplitude death phase in mean-field oscillator models.

\section{Complex Ginzburg-Landau Equation with driving}\label{sec:CGL}
\subsection{Model introduction}
The standard complex Ginzburg-Landau equation (CGLE),
\begin{equation}\label{eq:basicCGLE}
    \dot{A} = (1+i\omega_0)A-(1+i\alpha)|A|^2A+(1+i\beta)\nabla^2A,
\end{equation}
arises from general reaction diffusion equations as the medium undergoes a supercritical Hopf bifurcation \cite{newell1974envelope, kuramoto2003chemical}, and can be derived by an amplitude equation approach \cite{cross1993pattern, vanHecke1994Amplitude}, or more generally, as a low order approximation using a center manifold method \cite{kuramoto2003chemical}. This system oscillates at frequency $\omega_0-\alpha$. The parameter $\alpha$ induces phase-amplitude coupling and controls the nonlinear dispersion, while $\beta$ controls the linear dispersion \cite{aranson2002world}. The CGLE exhibits a gauge invariance of the type $A\to Ae^{\Phi}$, which can be used to eliminate the factor of $\omega_0$, indicating that there are no intrinsic timescales associated with linear response at small amplitudes in the absence of an external drive. We retain $\omega_0$ here since this gauge symmetry is  broken once a drive is added. Because of its generic character the CGLE has been applied to a very diverse set of systems, including biological and reaction diffusion systems, superfluidity and Bose-Einstein condensation, liquid crystals, and even string theory \cite{pismen1999vortices,aranson2002world}. It is valid under a broad range of conditions \cite{aranson2002world}, and has been suggested to be valid far from the onset of oscillations in some cases \cite{ouyang1996transition, leweke1994model}. More detailed information can be found the following reviews \cite{aranson2002world, garcia2012complex}.

In order to study how this oscillating medium can be controlled we introduce a drive that is harmonic in time and space:

\begin{align}\label{eq:DCGLE}
    \dot{A} &= (1+i\omega_0)A-(1+i\alpha)|A|^2A+(1+i\beta)\nabla^2A\\
    &\quad+Be^{i(-(\omega_0-\nu)t+\vec{k}_0\cdot\vec{x})}. \notag
\end{align}

Here $\nu$ is the difference between the natural frequency $\omega_0$ and the drive frequency, and is known as the ``detuning frequency", and $k_0$ is the wave number of the spatially nonuniform drive. In the absence of drive the CGLE supports plane wave solutions with the dispersion relation $\omega_{pw} = -\omega_0+\alpha+k_{pw}^2(\beta-\alpha)$, which are linearly stable only if $1+\alpha\beta>0$ and $k_{pw}$ is low enough. The condition on $\alpha$ and $\beta$ is known as the Benjamin-Feir-Newell criterion and for $1+\alpha\beta<0$ turbulent behavior is observed, which has rich behavior and especially when exposed to an external drive \cite{chate1999forcing}. In order to focus on more experimentally tractable questions we focus on the non-turbulent parameter regime. We also do not study drive wavenumbers $k_0$ close to one. One reason is formal, as plane waves in the CGLE with high wave numbers are unstable. Another is practical, $k_0 \to 1$ corresponds to a wavelength approaching the diffusion length and wavelengths of that scale are difficult to create and control for system sizes at the scale of many spatial oscillator experiments and the results would be similarly difficult to observe directly. 

\subsection{Analytic results}
Beginning with the spatially driven CGLE, Eq. \ref{eq:DCGLE} above, we boost to a gauge co-rotating with drive frequency, transforming $A\to Ae^{i(\omega_0-\nu)t}$. Carrying this out eliminates the factors of $\omega_0$:
\begin{equation}\label{eq:sCGLE}
    \dot{A} = (1+i\nu)A-(1+i\alpha)|A|^2A+(1+i\beta)\nabla^2A+Be^{i\vec{k}_0\cdot\vec{x}}
\end{equation}
In this gauge $A$ is synchronized to the drive when it is constant in time, so looking for synchronized solutions is equivalent to looking for fixed points of equation \ref{eq:sCGLE}. This can be further simplified by rewriting this equation as two real equations for the amplitude and phase of $A = Re^{i\Phi}$, and changing the gauge again, this time in space, as $\Phi\to \Phi-\vec{k}_0\cdot\vec{x}$, giving:
\begin{align} 
    \dot{R} &= R-R^3+\nabla^2R-R|\nabla\tilde{\Phi}-k_0|^2 \label{eq:R_simple}\\
    &\quad-\beta(2(\nabla R)\cdot(\nabla\tilde{\Phi}-k_0)+R\nabla^2\tilde{\Phi})+B\cos(\tilde{\Phi}) \notag \\
    R\dot{\tilde{\Phi}} &= \nu R-\alpha R^3+2(\nabla R)\cdot(\nabla\tilde{\Phi}+k_0)+R\nabla^2\tilde{\Phi} \label{eq:Phi_simple}\\
    &\quad+ \beta(\nabla^2R-R|\nabla\tilde{\Phi}+k_0|^2)-B\sin(\tilde{\Phi}).\notag
\end{align}
The drive is now constant in this gauge so we ignore spatial variations in $R$ and $\Phi$, giving:
\begin{align}
    \dot{R}&=R(1-k_0^2)-R^3+B\cos(\Phi) \label{eq:simpleODEsR}\\
    R\dot{\tilde{\Phi}}&=R(\nu-\beta k_0^2)-\alpha R^3-B\sin(\Phi). \label{eq:simpleODEsPhi}
\end{align}

Synchronized solutions are fixed points of these equations. Taking the fixed point condition and eliminating $\Phi$ we arrive at an expression for $B^2$ as a function of the fixed point amplitude $R_0$:
\begin{equation}\label{eq:sCGLE_fxpt}
    B^2=R^2((R^2-1+k_0^2)^2+(\nu-\beta k_0^2-\alpha R^2)^2).
\end{equation}

In the limit of no drive, where $B\to0$, $k_0\to0$, and $\nu\to\alpha$, Eq. \ref{eq:sCGLE_fxpt} (considered as a cubic equation in $R^2$) has a double root at $R^2 = 1$, generalizing a previous results for uniform drive \cite{chate1999forcing}. For small $B$ this double root can turn into two real roots, one of which is stable and represents the synchronized solution. Therefore we can estimate the boundary of the synchronized region, known as an Arnol'd tongue, by requiring that there exist exactly two coinciding positive real solutions. This is equivalent to requiring that the discriminant of Eq. \ref{eq:sCGLE_fxpt}, considered as a cubic equation in $R^2$, vanishes. If we do so and identify $\epsilon = \nu-\alpha+k_0^2(\alpha-\beta)$ we find a simple expression for the the Arnol'd tongue that approximates Eq. \ref{eq:sCGLE_fxpt} for small $\epsilon$:
\begin{equation}\label{eq:ATwidth}
    B^2 = \frac{\epsilon^2(1-k_0^2)}{1+\alpha^2}.
\end{equation}

This is a generalization of the result for uniform driving found in \cite{chate1999forcing}, and allows an estimation of the range of frequencies that a drive of a given amplitude can synchronize, without needing to directly calculate the linear stability of the full system. This expression has a notable feature: the center of the Arnol'd tongue, $\nu^* = \alpha + k_0^2(\alpha-\beta)$ depends on $k_0$, the wavenumber of the drive. This means that the range of frequencies that can be synchronized to the drive can itself be controlled by the spatial distribution of the drive.

\subsection{Periodic boundary conditions}
In order to validate the analytic results it is necessary to perform simulations, which introduces complications not present in the preceding analysis. In particular it is necessary to choose the boundary conditions for the full PDE since the preceding Aronl'd tongue analysis did not make any assumptions about them. For systems where one does not expect strong boundary effects or where one is interested primarily in behavior far from the boundaries it is often practical to use periodic boundary conditions. These are especially natural conditions for oscillatory phenomena, and therefore we test the Arnol'd tongue described by Eq. \ref{eq:ATwidth} first with simulations using periodic boundary conditions. 

\begin{figure}[h!]
    \centering
     \includegraphics[width=0.45\textwidth]{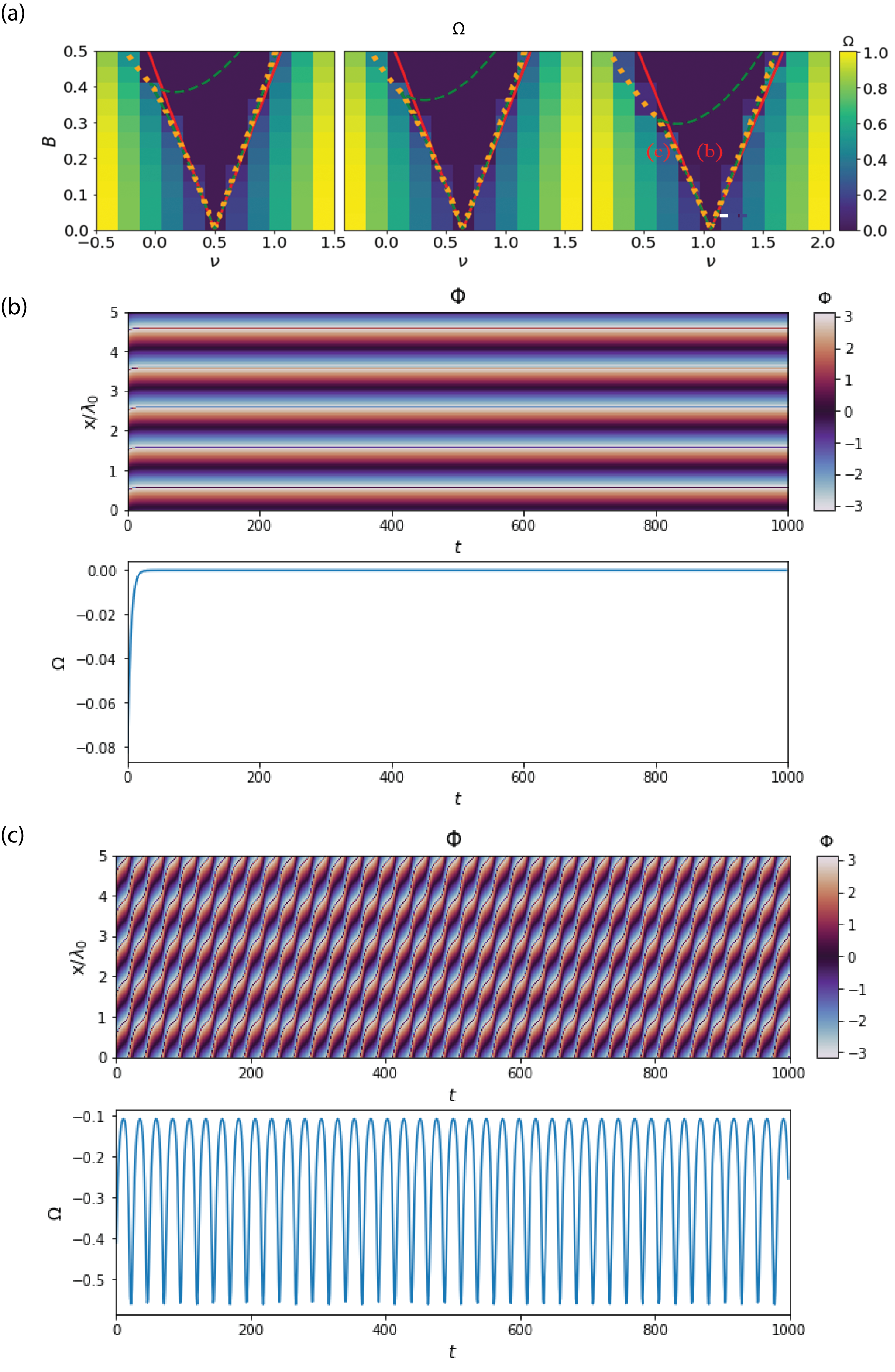}
         
    \hfill
    \caption{{\bf Changing the wavenumber of the drive shifts the Arnol'd tongue}. {\bf (a)} Plots of $\Omega = \langle\partial_t\Phi\rangle$ (the spatially-averaged time derivative of the phase) as a function of the drive parameters $\nu$, the detuning frequency (the difference between the drive frequency and the natural frequency), and $B$, the drive amplitude. The drive wavenumber $k_0$ increases from left to right, taking the values 0, 0.2, and 0.4, respectively. Red contours show the approximate Arnol'd tongue described by Eq. \ref{eq:ATwidth}, which estimates the region in which the medium is synchronized to the drive. The dotted orange contours show the boundary of the region in which the ODEs describing a uniform system, Eqs. \ref{eq:simpleODEsR} \& \ref{eq:simpleODEsPhi}, are stably synchronized to the drive. The dashed green contour shows the region within which the ODE system has exactly three real solutions for $R_0^2$. {\bf (b)} $\Phi$ and $\Omega$ as a function of space and/or time at the drive parameters indicated in (a), $\nu = 1.06$, $B = 0.2$.{\bf (c)} Same as (b) but for the drive parameters marked by the (c) in (a), $\nu = 0.7$, $B = 0.2$. Distances are measured in units of the drive wavelength: $\lambda_0 = 2*\pi/k_0$. Parameters: $\alpha = 0.5$ and $\beta = 4$.} 
    \label{fig:SimpleATs}
\end{figure}

Arnol'd tongues for several values of $k_0$ are plotted in Fig. \ref{fig:SimpleATs}a, along with time-averaged values of the quantity $\Omega = \langle\partial_t\Phi\rangle$, with angle brackets indicating a spatial and temporal average. The shift in the range of frequencies that can be synchronized shift noticeably as a function of $k_0$. Since the system is observed in the gauge co-rotating with the drive frequency solutions that are synchronized with the drive will have $\Omega = 0$ and $\Omega \neq 0$ indicates non-synchronized behavior. This is shown in Fig. \ref{fig:SimpleATs}b, c where plots of $\Omega$ and the phase $\Phi$ are shown for a system of size $L = 12\pi/k_0$, in a gauge co-rotating with the drive frequency but not with the drive wavenumber to show the spatial nonuniformity. The simulations agree very well with the analytic results, as might be expected considering that periodic boundary conditions are a natural choice for solutions that are themselves periodic in space. 

It is also possible to extract information about the bifurcation structure of this system from Fig. \ref{fig:SimpleATs}a. In addition to the approximate Arnol'd tongue Eq. \ref{eq:ATwidth}, we plot the boundaries of the region of stability of the fixed point of the ODE system describing the uniform solution,Eqs. \ref{eq:simpleODEsR} \& \ref{eq:simpleODEsPhi}, as well as the region where these ODEs have exactly 3 fixed points. The structure of the boundaries of these regions suggests that the type of bifurcation between synchronized and desynchronized solutions changes depending on the drive amplitude $B$. At low $B$ the bifurcation appears to be of the SNIPER (saddle-node infinite period) type since the number of fixed points changes from 1 to 3. At higher $B$ the number of fixed points remains constant at 1, indicating a Hopf bifurcation. These two synchronized regions are separated by a boundary where the number of fixed points changes, suggesting a standard saddle-node bifurcation.

A structure of this type, with both SNIPER and Hopf bifurcations connected by a saddle-node bifurcation has been shown to exist for the driven Kuramoto model with oscillators with heterogeneous freqeuencies \cite{antonsen2008external, childs2008stability}. In this mean-field model it has been analytically shown that two dynamically distinct regions of synchronized solutions exist, with different numbers of fixed points. In the driven Kuramoto model these regions are separated by a saddle-node bifurcation in a large region of parameter space, and a much more complex series of bifurcations close to the point where the two different synchronized regions meet the desynchronized region. This correspondence between the driven CGLE and driven Kuramoto model is surprising at first glance since the CGLE is a model of a system with spatial extent while the Kuramoto model is a mean-field system. This suggests that this behavior may be generic to driven coupled-oscillator models, but much more work is needed to further tease out this connection.

\subsection{Zero-flux boundary conditions}\label{sec:CGL0flux}
\begin{figure}[htb]
    \centering
    \includegraphics[width=0.4\textwidth]{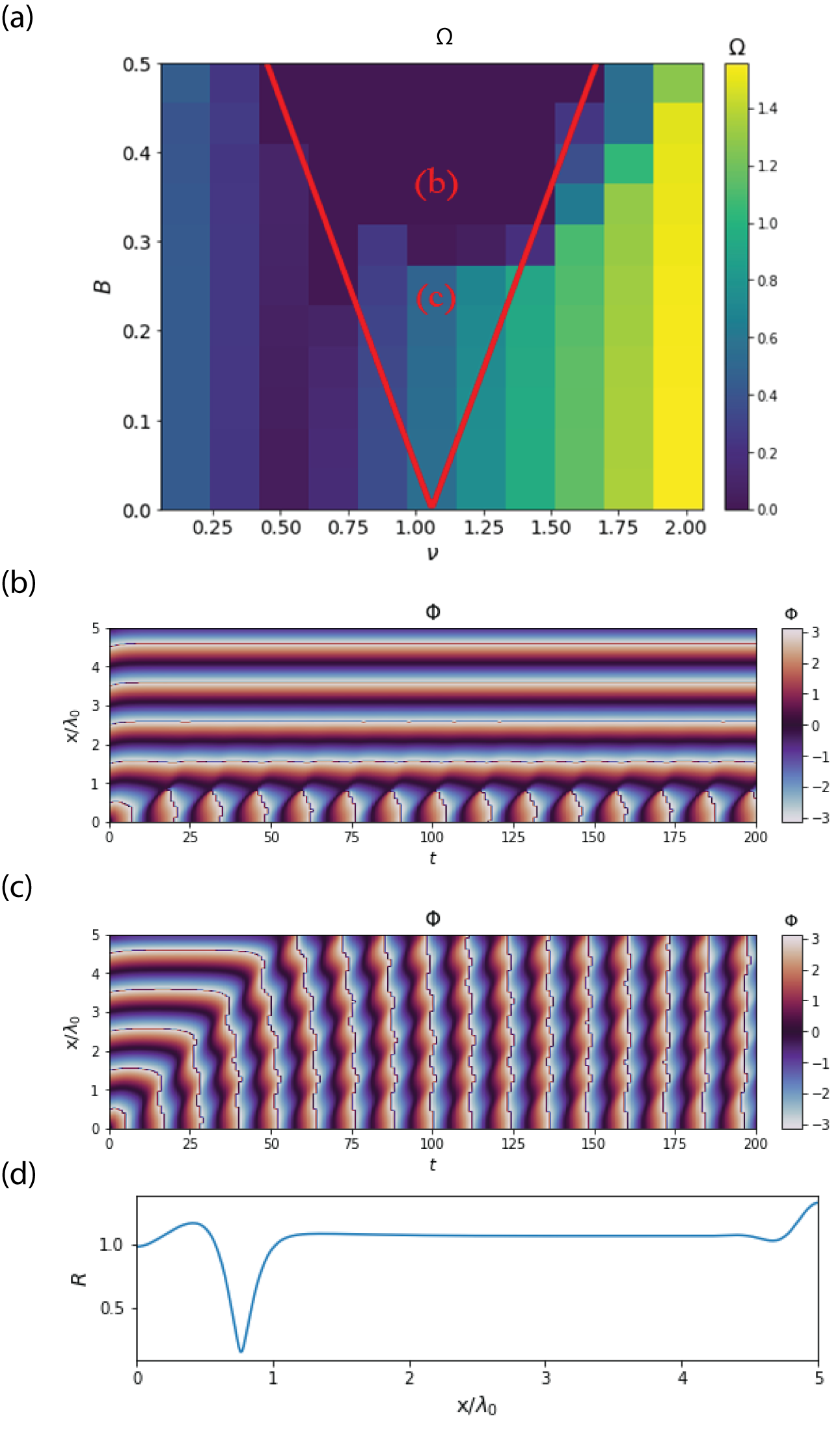}
    \caption{{\bf Zero flux boundary conditions can cause desynchronization in directly coupled CGL systems even where analytical calcualtions predict synchronization}.{\bf (a)} Analytically calculated boundary of Arnol'd tongue for $k_0 = 0.4$ (red) using Eq. \ref{eq:ATwidth} and numerical simulations of $\Omega = \langle\partial_t\Phi\rangle$ as a function of the detuning frequency $\nu$ and driving amplitude $B$. {\bf (b)} The phase $\Phi$ for the point labeled (b) in (a), $\nu = 1.06$, $B = 0.2$. {\bf (c)} The phase $\Phi$ for the point labeled (c) in (a), $\nu = 1.06$, $B = 0.35$. {\bf (d)} The amplitude $R$ as a function of spatial position $x$  at $t=100$ for the parameter set shown in (b).
     Parameters: $\alpha = 0.5$, $\beta = 4$, and $k_0 = 0.4$.}
    \label{fig:0FluxAT}
\end{figure}

Boundary conditions play an important role in the behavior of PDEs generally, but the analytic results we presented were agnostic of them. The results were valid for periodic boundary conditions which are often practical for bridging simulation and theory, but other types of boundary conditions are more appropriate for many experimental applications. For applications involving diffusing media it is often the case that zero-flux boundary conditions are the most appropriate for systems of finite size. Here we show that these boundary conditions can have a strong effect on the synchronization, especially close to the border of the Arnol'd tongue. This is shown in Fig. \ref{fig:0FluxAT}a, which takes the same medium parameters as Fig. \ref{fig:SimpleATs}a but with zero-flux boundary conditions as opposed to periodic boundary conditions. 

The system was simulated with the same parameters as Fig. \ref{fig:SimpleATs} for $k_0 = 0.4$ and the results are shown in Fig \ref{fig:0FluxAT}. At low forcing amplitude $B$ the area within the Arnol'd tongue shows nonzero $\Omega$, indicating desynchronzation, shown in Fig. \ref{fig:0FluxAT}b. The system appears to be stably synchronized to the drive for early times, at least in the bulk of the system, but a shock-like effect originates at the $x = 0$ boundary and travels across the system, desynchronizing it. Even at higher forcing parameters where the steady state value of $\Omega$ becomes small there is still a boundary effect, shown in Fig. \ref{fig:0FluxAT}c, in which there is a region close to one of the boundaries that is desynchronized even though the bulk of the system is synchronized to the drive. Such regions in solutions of the CGLE are known as phase reconnections and are accompanied by the amplitude vanishing locally, shown in Fig. \ref{fig:0FluxAT}d. While the boundary of the synchronized region appears similar to the boundary of the region where the number of fixed points changes from 1 to 3, the desynchronization involves a highly spatially organized process, which cannot be accounted for by the ODE model.

Additionally, there are regions outside the Arnol'd tongue with $\Omega \sim 0$, indicating that they are synchronized with the drive (i.e.  there is a 1:1 coupling between observed frequency and drive frequency). However, in these regions we find a rich synchronization structure in \emph{space}, with the spatial wavelengths of both the phase and amplitude related to spatial wavelength of the external drive through the ratio of two integers. This non-trivial ``spatial'' phase locking occurs despite the fact that the oscillators are temporally synchronized with the drive. While it is well understood that Arnol'd tongues can be surrounded by other tongues for other synchronization ratios \cite{garcia2008high}, this phenomenon is different since each individual oscillator in the medium is still 1:1 phase locked to the drive in time. Instead, the higher order synchronization is in space

\begin{figure}[h!]
    \centering

     \includegraphics[width=0.45\textwidth]{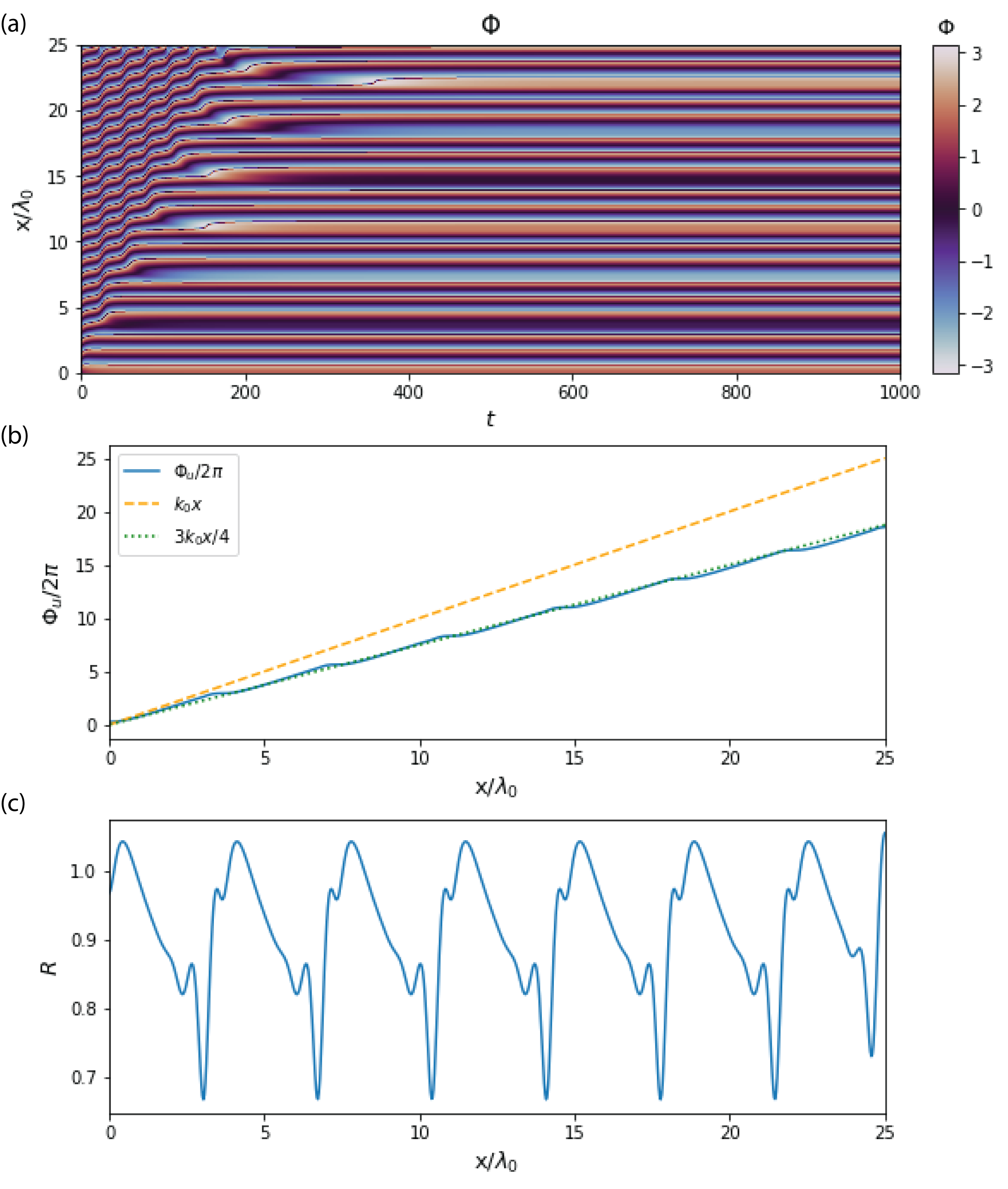}

    \caption{{\bf Zero flux boundary conditions can lead to more complex spatial mode-locking patterns in regions: 4:3 spatial locking}. {\bf (a)} The phase $\Phi$ as a function of space and time for zero-flux boundary conditions. {\bf (b)} The winding number, defined in terms of the unwrapped phase $\Phi_u$, as $\Phi_u/2\pi$, as a function of $x$ to $t = 900$, with the dashed line showing 1:1 spatial locking and the dotted line 4:3 spatial locking.{\bf (c)} The amplitude $R$ as a function of spatial position $x$ at $t = 900$. Parameters:$\alpha = 0.5$, $\beta = 4$, $k_0 = 0.4$, $B = 0.25$, and $\nu = 0.66$}
    \label{fig:0Flux4_to_3}
\end{figure}

\begin{figure}[h!]
    \centering
     \includegraphics[width=0.45\textwidth]{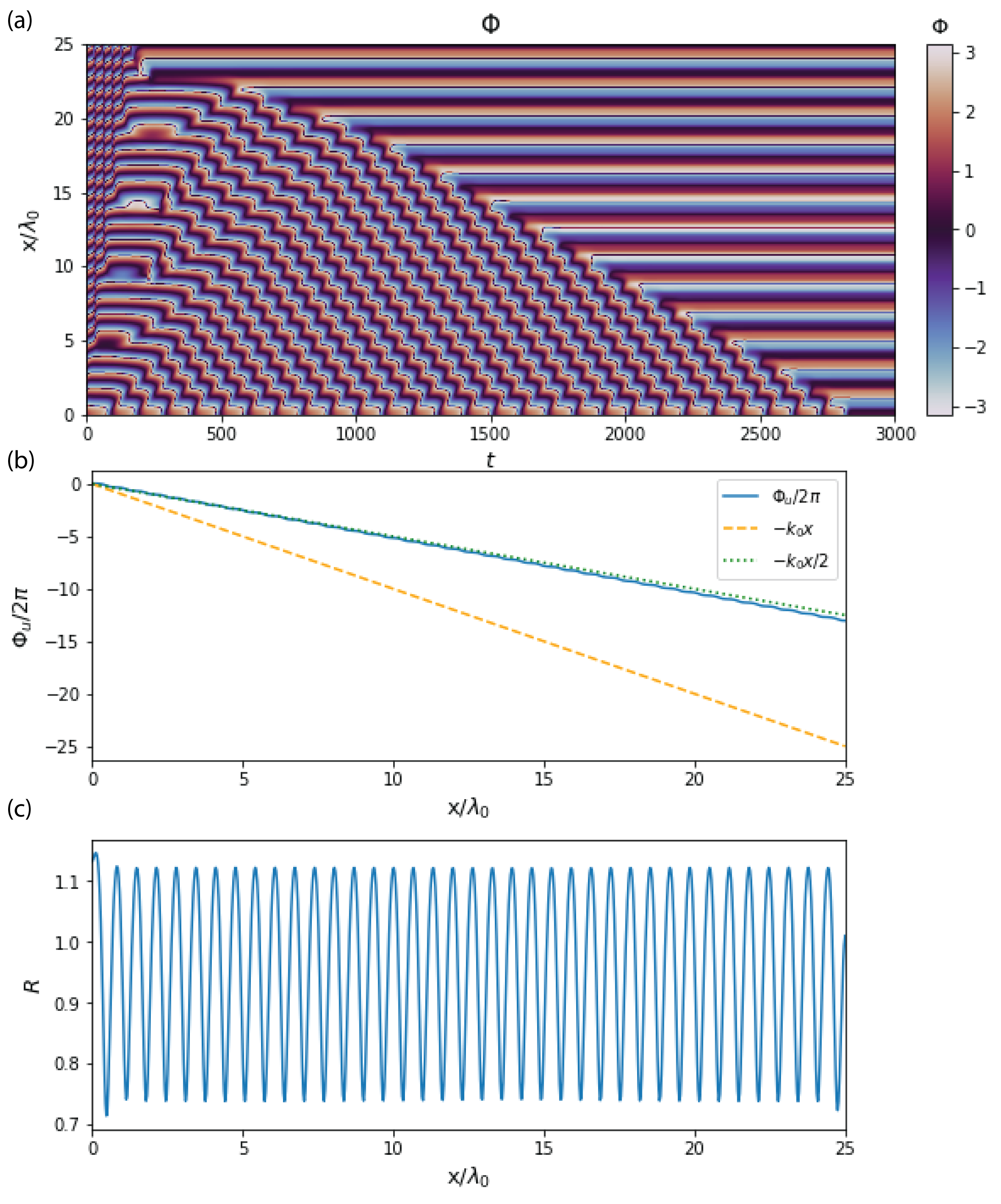}

    \caption{{\bf Zero flux boundary conditions can lead to more complex spatial mode-locking patterns in regions: 2:1 spatial locking}. {\bf (a)} The phase $\Phi$ as a function of space and time for zero-flux boundary conditions. {\bf (b)} The winding number, defined in terms of the unwrapped phase $\Phi_u$, as $\Phi_u/2\pi$, as a function of $x$ to $t = 2900$, with the dashed line showing 1:1 spatial locking and the dotted line 2:1 spatial locking.{\bf (c)} The amplitude $R$ as a function of spatial position $x$ at $t = 2900$. Parameters:$\alpha = 0.5$, $\beta = 4$, $k_0 = 0.4$, $B = 0.25$, and $\nu = 0.68$}
    \label{fig:0Flux2_to_1}
\end{figure}

There appear to be a rich variety of types of such synchronized solutions, two of which can be seen in Figs. \ref{fig:0Flux2_to_1}  and \ref{fig:0Flux4_to_3}. For the same value of drive amplitude $B$ two very similar choices for detuning frequency produce dramatically different patterns. For $\nu = 0.66$ (Fig. \ref{fig:0Flux4_to_3}) we find a region that shows an approximate 4:3 ``spatial phase locking'' in space (i.e. the spatial wavelength of the phase
is three-fourths the spatial wavelength of the external drive) in addition to the usual 1:1 phase locking in time. This can be seen more clearly by observing that at a fixed time, the phase increase in space (Fig. \ref{fig:0Flux4_to_3}b). Additionally, there is a highly nontrivial phase-amplitude coupling which can be seen by noting the highly irregular pattern formed by the amplitude $R$ as a function of space (\ref{fig:0Flux4_to_3}c) .  However, despite this irregularity, the amplitude still has a well defined wavelength characterizing its spatial modulation.  The spatial wavelength of $R$ differs from both the wavelength characterizing the spatial modulation of the phase and the wavelength describing the external drive. This can be seen by noting that whereas the amplitude $R$ completes between 6 and 7 spatial oscillations traversing the system, the phase completes approximately 18 and the drive completes 25.

Slightly increasing the detuning frequency $\nu$ to 0.68 induces a different pattern (Fig. \ref{fig:0Flux2_to_1}a). Initially, the pattern appears to desynchronize by a similar traveling shockwave mechanism as shown in Fig. \ref{fig:0FluxAT}, but at later times the system enters an approximate 2:1 ``spatial phase locking'' where the spatial wavelength of phase is half of the wavelength of the external drive (i.e. the phase oscillates twice as fast in space as the external drive). This can be clearly seen in Fig. \ref{fig:0Flux2_to_1}b. It also shows a nontrivial phase-amplitude coupling,  with the amplitude completing 36 spatial oscillations while transversing the system compared to the drive which completes only 24 (Fig. \ref{fig:0Flux2_to_1}c). 

These complex behaviors are found very close together, near the point where the boundary between the SNIPER, Hopf, and saddle-node bifurcations all intersect. In the driven Kuramoto model this region is known to have a quite complex bifurcation structure, including a codimension-2 Takens-Bogdanov point where an additional homoclinic bifurcation can be found \cite{childs2008stability}. While we do not reproduce a full bifucation analysis of the CGLE here, the complexity of this known system suggests that the complex synchronization structure we observe here is tied to a rich, local bifurcation structure in this region of the $\nu-B$ plane.

\section{Driven oscillators with external medium}\label{sec:emCGL}
\subsection{Model introduction}
In biological applications such as quorum sensing the components of the chemical oscillators are contained within individual cells and are usually confined within the cells. Instead the cells produce a signaling molecule which diffuses or is excreted out of the cell and acts as an external proxy for the local population's internal oscillation state. 
In the case of \emph{Dictyostelium} collective oscillations this extracellular signal has been identified as cyclic AMP \cite{mcmains2008oscillatory}. 
It is this extracellular medium that diffuses between the cells and couples the cellular population. This leads to a nonlocal coupling of local oscillators which is not equivalent to the standard CGLE, even close to the onset of oscillations \cite{tanaka2003complex,kuramoto1997phase}, and exhibits behaviors that are qualitatively distinct from the standard CGLE \cite{tanaka2003complex}. This system is described by the equations:
\begin{align}\label{eq:basicEMCGLE}
    \dot{A} &= (1+i\omega_0)A-(1+i\alpha)|A|^2A-D(A-Z)\\
    \dot{Z} &= \rho D(A-Z)+(1+i\beta)\nabla^2Z-JZ,
\end{align}
and we will refer to this model as the ``external medium complex Ginzbug-Landau equation" (emCGLE). 
The parameter $J$ is the degradation rate of the external medium, $D$ is a coupling constant between the external medium and the oscillating field, and $\rho$ is the density of the oscillating field $A$. The addition of an external medium $Z$ that does not naturally oscillate induces ``memory" effects in the form of additional timescales that cannot be transformed away. 

We add the drive term to the oscillator field $A$ as before. This can be understood biologically as using a mechanism such as an optogenetic control of the individual cells to activate regulatory pathways that influence the biological oscillator, inducing increases in internal cAMP that are released into the external medium and driving other cells through this medium. The emCGLE with drive then takes the following form:

\begin{align}\label{eq:emCGLEdriven}
    \dot{A} &= (1+i\omega_0)A-(1+i\alpha)|A|^2A-D(A-Z)\\
    &\quad+Be^{i(-(\omega_0-\nu)t+\vec{k}_0\cdot\vec{x})}. \notag\\
    \dot{Z} &= \rho D(A-Z)+(1+i\beta)\nabla^2Z-JZ
\end{align}

Note that by introducing an external coupling medium with its own dynamics the emCGLE introduces a number of additional timescales that now can affect the behavior. First, the addition of the external medium $Z$ has broken the gauge symmetry of the CGLE so the factor of $\omega_0$ cannot be eliminated. Instead, the rate (inverse timescale) associated with the frequency mismatch $\delta\omega = \omega_0-\nu$ is absorbed into the dynamics of the external medium $Z$. The external medium field $Z$ also introduces two other rates:  $\omega_{Zr} = J+\rho D$, the rate associated with relaxation of the medium to small perturbations, as well as $\omega_{Zc} = \rho D$, the rate associated with coupling the oscillators and the medium . 

The relations between these timescales can provide intuition about the dynamics of the emCGLE in a given parameter regime. If we have $\omega_{Zr}\gg1$ (e.g., if the signaling molecule is degraded quickly) then the medium will be pinned to zero. On the other hand, if $\omega_{Zc}\ll1$ the medium responds almost instantly to the oscillator and in this limit the original CGLE is recovered. The emCGLE also introduces corresponding rates for the oscillator field $A$. There is a relaxation rate $\omega_{Ar} = 1-D$ and a coupling rate $\omega_{Ac} = D$. While these are not independent, we distinguish them to note that the form of $\omega_{Ar}$ suggests the possibility of some kind of change in behavior at $D=1$, which we discuss below.

A mean-field version of this model without spatial structure has been studied previously \cite{schwab2012dynamical, schwab2012kuramoto} and has been shown to possess a transition from a collectively oscillating state to an ``amplitude death" state in which delays induced by the medium can cause individual oscillators to cease oscillating. This can come about in several ways. It can be caused by the external medium being degraded to the point where it can no longer maintain oscillations and couple the oscillators, or by a ``dynamic" mechanism in which the oscillation is too fast for the medium to keep up. Here we probe how this state can be accessed and synchronized to a drive.

\subsection{Analytic results and simulations}
The emCGLE can be understood using the same methods that were used to study the standard CGLE. First we examine the uniform fixed points of the dynamics. To do so we transform to the gauge in which the drive is constant, taking $A\to Ae^{i((\omega_0-\nu)t-k_0x)}$.  This results in the following equations:
\begin{align}
    \dot{{A}} &= (1+i\nu){A}+(1+i\alpha)|{A}|^2{A}-D({A}-{Z})+B\\
    \dot{{Z}} &= \rho D({A}-{Z})+(1+i\beta)\left(\nabla^2{Z}+2ik_0\cdot\nabla{Z}\right) \nonumber \\
    &\quad-((J+k_0^2)+i(\omega_0-\nu+\beta k_0^2)){Z}
\end{align}
In this gauge it can be seen that, for the purposes of finding uniform solutions, changing the wave number of the drive is equivalent to a suitable shift in the ``memory parameters" $J$ and $\omega_0$ (except for questions of linear stability which we do not analytically address in this paper). Therefore we set $k_0=0$ without loss of generality for purposes of calculating the Arnol'd tongue. We proceed as before by rewriting the model in terms of amplitude and phase variables: $A = R e^{i\Phi}$ and $Z = r e^{i\theta}$. Then, assuming a spatially uniform solution we arrive at the following system of equations:
\begin{align}
    \dot{R} &= (1-D)R-R^3+Dr\cos(\theta-\Phi)+B\cos(\Phi) \label{eq:emCGLERdot}\\
    R\dot{\Phi} &= \nu R-\alpha R^3+Dr\sin(\theta-\Phi)-B\sin(\Phi) \label{eq:emCGLEPhidot}\\
    \dot{r} &= -\rho D r +\rho DR\cos(\Phi-\theta)-Jr\\
    r\dot{\theta} &= \rho DR\sin(\Phi-\theta)-(\omega_0-\nu) r.
\end{align}

The fixed points of these equations represent phase-locked solutions. The last two equations can be combined to eliminate the $\sin(\theta-\phi)$ and $\cos(\theta-\phi)$ terms. This also gives the relation $q R = r$, where $q = \sqrt{\frac{(\rho D)^2}{(\rho D+J)^2+(\omega_0-\nu)^2}}$. The $\sin(\Phi)$ and $\cos(\Phi)$ terms can then be eliminated just as for the standard CGLE, giving the following expression:
\begin{align}\label{eq:discexpr}
\begin{split}
    B^2 = &R^2\bigg[\left((D-1)+R^2-Dq^2\left(\frac{J+\rho D}{\rho D}\right)\right)^2\\
    &\quad+\left(\nu-\alpha R^2 - Dq^2\frac{\omega_0-\nu}{\rho D}\right)^2\bigg].
\end{split}
\end{align}
This is a cubic equation in $R^2$, analogous to Eq. \ref{eq:sCGLE_fxpt}. 


From Eq. \ref{eq:discexpr} we can follow a similar procedure as for the CGLE and calculate the discriminant. Once this is accomplished We can find the center of the Arnol'd tongue $\nu^*$ by solving for $\nu$ when $B=0$. This condition reduces to an implicit equation for $\nu^*$:
\begin{equation}
    \nu^* = \alpha\left(1+D\left(q(\nu^*)^2\frac{J+\rho D}{\rho D}-1\right)\right)+q(\nu^*)^2D\frac{\delta\omega(\nu^*)}{\rho D}.
\end{equation}

Taking $\epsilon = \nu-\nu^*$ to be small and solving for the condition that the discriminant vanishes for small $B$ we arrive at an expression for the Arnol'd tongue:

\begin{equation}\label{eq:MediumAT}
     B_{J\omega}^2 = \frac{|1+D(q^2\hat{J}-1)|}{1+\alpha^2}\left(\nu-\alpha(1+D(q^2\hat{J}-1))-q^2D\delta\hat{\omega}\right)^2
\end{equation}
where we have suppressed the $\nu$ dependence in $q$ and $\delta\omega$ and renormalized the memory coefficients $\hat{J} = 1+J/\rho D$ and $\delta\hat{\omega} = (\omega_0-\nu)/\rho D$. 

Equation \ref{eq:MediumAT} is richer than the Arnol'd tongue expression for the model without the extracellular medium. It is possible to recover Eq. \ref{eq:ATwidth} for $k_0 = 0$ from Eq. \ref{eq:MediumAT} in the limit of high density, $\rho \to \infty$. The reason $k_0$ does not appear in this case is that $k_0$ enters as shifts in the memory parameters and all the memory parameters are effectively rescaled to zero in the limit of large density. 

Some intuition can be gained from this theory by examining how it is affected by the different timescales. First, we consider the limit in which $Z$ responds infinitely quickly to $A$, $\omega_{Zc}\to\infty$. In this limit we have $Z = A$, recovering the CGLE. This limit can also be understood intuitively as the limit of large density. 

In order to understand the opposite limit we can interpret the factor $q$ as a ratio of the timescales associated with the medium: $q = \sqrt{\omega_{Zc}^2/ (\omega_{Zr}^2 + \delta\omega^2})$. Defining a complex ``memory loss" rate as $\omega_m = \omega_{Zr}+i\delta\omega$ which combines the effects of degradation of the medium and the rate of rotation relative to the external medium. We can then rewrite the expression for $q$ as $q = |\omega_{Zc}/\omega_m|$, showing that as memory becomes short compared to the coupling response time, $\omega_m\gg\omega_{Zc}$, then $r$ becomes small. When this occurs coupling between individual oscillators is lost and the only interaction is between the drive and the individual local oscillators. To see how this affects the process of synchronization, observe that if $D>1$ then Eq. \ref{eq:emCGLERdot} implies that for $r=0$ we also have $R=0$ in the absence of a drive. In the study of the synchronization of systems of coupled oscillators without a drive this phenomenon is know as ``amplitude death", and has been studied in mean-field versions of this system \cite{schwab2012dynamical, schwab2012kuramoto}. Amplitude death is a response characteristic of coupled oscillator systems with some kind of delay in the coupling \cite{strogatz1998death}. In this case the delay is induced by the ``memory" effects of the medium.

\begin{figure*}[h!]
    \centering
    \includegraphics[width=0.9\textwidth]{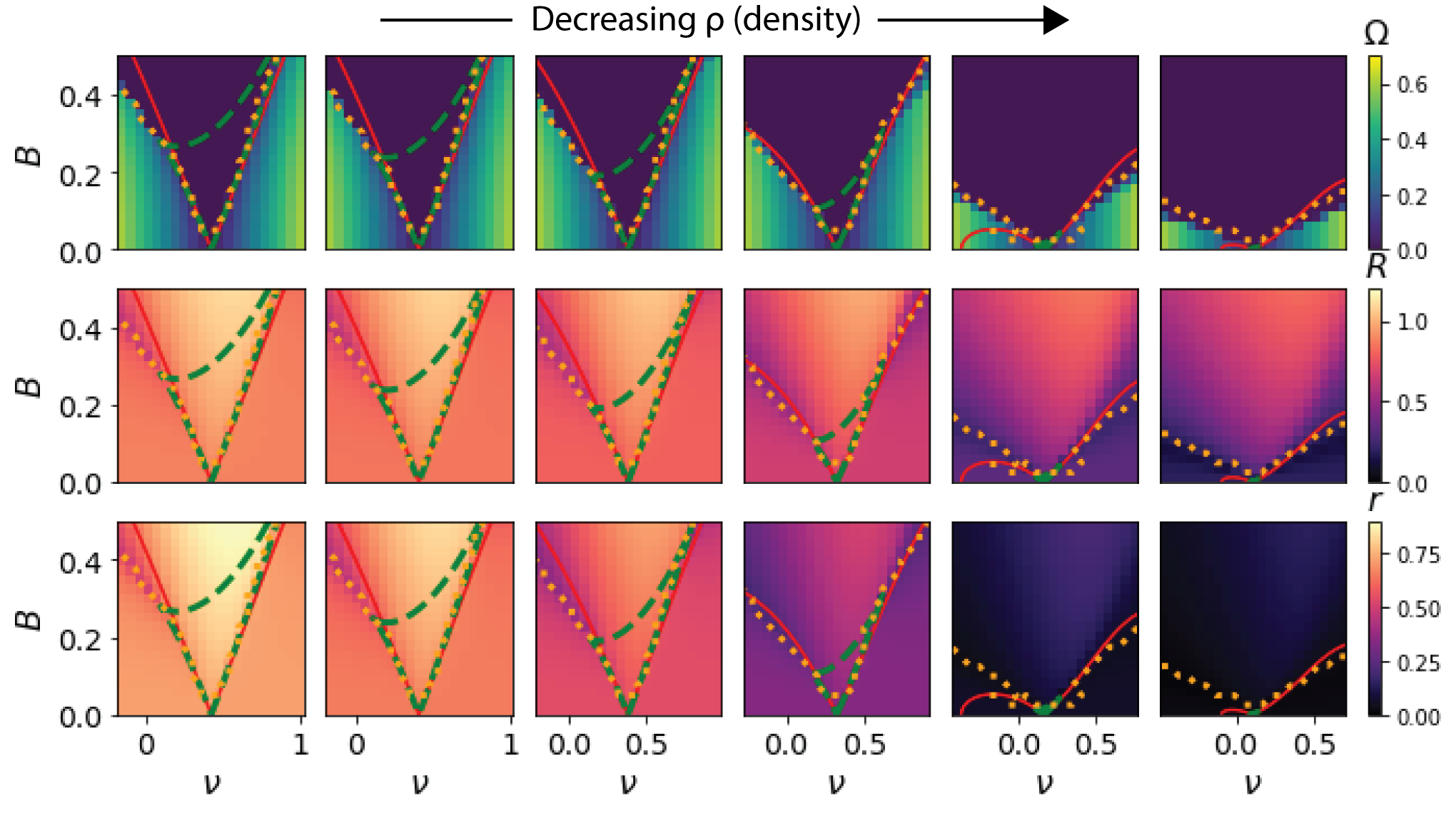}
    \caption{{\bf CGL Model coupled through an external medium exhibits a low density amplitude death regime in which very wide ranges of frequencies can be synchronized with small driving amplitudes}. {\bf (Top)} $\Omega$, {\bf (Middle)} $R$, and {\bf (Bottom)} $r$ as a function of $\nu$ and $B$, with analytic boundaries of the approximate Arnol'd tongue, Eq. \ref{eq:MediumAT} (red solid contour), the existence of a stable, synchronized fixed point (orange dotted contour) and the are within which there are exactly three synchronized fixed points (green dashed contour). This indicates the same approximate bifurcation structure as in the driven Kuramaoto model and the CGLE, with SNIPER, saddle-node, and Hopf bifurcations. As the density declines the coupling to the medium happens on a slower and slower rate compared to the rates associated with medium turnover ($J$) or the drive frequency ($\omega_0-\nu$). As this happens the medium has effectively zero amplitude (bottom right) and no longer couples the individual oscillators. Then, either they oscillate along with the drive or they also decay to a low amplitude. In this limit, the width of the Arnol'd tongue increases and becomes proportional to $B^2/(D-1)$. Parameters: $\alpha = 0.5$, $\beta = 1$, $k_0 = 0.1$, $D = 1.1$, $J = 0.5$, $\omega_0 = 0.5$, and $\rho= 2.0, 1.5, 1.0, 0.5, 0.15,0.1$ from left to right. Note that there are no orange lines in right most plots since pre-factor in Eq. \ref{eq:MediumAT} is zero.}
    \label{fig:mediumATs_disc}
\end{figure*}

To study the effects of a drive on the low density phase we consider the case where $D>1$. When $\rho\to0$ and $D>1$ the oscillatory medium $A$ will decay to zero without the influence of the drive. We can observe the onset of this process for nonzero drive in Fig. \ref{fig:mediumATs_disc} in which the density is gradually decreased. Most obviously, the Arnol'd tongue gets much wider as the density declines, enabling the medium to be synchronized by a much wider range of frequencies as the coupling through the medium becomes more difficult. 

As this occurs several things happen. The region in which there are 3 unique positive solutions for Eq. \ref{eq:discexpr}, which is enclosed in an orange dashed contour in Fig. \ref{fig:mediumATs_disc}, becomes much smaller, almost vanishing entirely for low densities. In the driven Kuramoto model it was found that the location of the saddle-node bifurcation scales with the steady-state amplitude \cite{childs2008stability}, and we see a similar decline in the steady state amplitude $R$ as this region shrinks. Furthermore, as the density declines and $q\to0$ we can see that the amplitude of the medium $r$ approaches zero, as expected, indicating that the medium $Z$ is no longer coupling the oscillator field $A$. In this low density regime the oscillators as well as the medium drop to low amplitude. Here Eq. \ref{eq:emCGLEPhidot} shows that when $q\to0$ the oscillator field $A$ will decay to 0 without the drive, and we see that when the system is not locked to the drive the oscillator amplitude $R$ drops to very low values, kept slightly above zero by the presence of the drive.

In summary, we see that at low-densities oscillators can easily synchronize to an external drive over a wide range of frequencies. This region corresponds to densities at which the corresponding mean-field model without space exhibits amplitude death. This suggests that it maybe easier to entrain coupled oscillator systems at low densities where normally collective oscillations are weak or even absent.


\section{Experimental implications of our results}

These results have exciting implications for controlling biological and synthetic oscillatory systems. There are clear experimental tests to check if the theoretical models studied here are representative of the experimental systems we seek to control. One clear prediction of our model is that they should be controllable using a wide range of drive frequencies if we can tune the experimental system close to the dynamic death phase. This can be accomplished by, for example, degrading the external medium. In particular, our models suggest that the amplitude of the external drive can be relatively small compared to the amplitude of the oscillations and still induce synchronization. This suggests that relatively small perturbations should still be able to change population behaviors.

Some previous experimental efforts \cite{sgro2015intracellular, gregor2010onset} have manipulated the external medium either through degradation alone or adding a uniform stimulus to the population through the fluid surrounding the cells. These different experimental manipulations of the medium tune coupling between individuals and their collective oscillation frequencies, but these new theoretical results suggest that one could also take advantage of the individual oscillators themselves to drive these types of cellular collectives, either alone or in conjunction with further manipulations to the external medium. 

Specifically, one could imagine using optogenetics to drive a subpopulation of individuals using a spatially-structure the external laser illumination. Optogenetic production of the molecule that cells use to signal one another should, in many systems, trigger release of this molecule to the external environment, mimicking the natural generation and release that couples the individual oscillators through the external medium.  If the individual cells are seeded densely enough, simply projecting the desired spatial drive pattern onto the cells should create the Arnol'd tongues observed in the results here, as well as the complex spatially mode-locked patterns. Crucially, these phenomena should be robust across a wide range of drive frequencies for the optogenetic control.  

\section{Discussion}

In this paper we study how generic models of oscillatory media can be controlled by spatially varying harmonic drives. We derived simple conditions that can be used to estimate when a drive of a given detuning frequency $\nu$ and amplitude $B$ will synchronize a medium in the CGLE. We showed that the spatial arrangement of the drive, in particular the wavelength, can control the temporal aspect of the synchronization process, in particular the frequency band that can be synchronized.

We compared analytically convenient periodic boundary condition with the more experimentally appropriate zero-flux boundary conditions in the CGLE and found that they can cause a dramatic effect on the observed synchronization phenomena. Regions that synchronize to the drive for periodic systems were destabilized by travelling shocks for zero-flux systems. Additionally, some regions that were unsynchronized for periodic systems showed temporal synchronization with complex spatial structure for zero-flux systems. 

We also investigated using a drive to control the emCGLE, since this model more closely mimics cell-based oscillations such as those in yeast glycolysis or \emph{Dictyostelium} aggregation. This system also showed a variable range of frequencies that can be locked to a drive, but the drive wavelength is absorbed into a complex ``memory loss" rate in the external medium that competes with a coupling rate from the oscillator field. When the coupling from the oscillators is weak compared to memory loss the external medium ceases to couple individual oscillators and they interact only with the drive. In this regime the system synchronizes to a much broader range of drive frequencies and the oscillator amplitude declines dramatically when it is not synchronized to the drive.

We demonstrated that the two models share a common bifurcation structure. For both models the region in the $\nu-B$ plane in which the system is synchronized to the drive is broken up into two components, one with three fixed points and one with only one. Furthermore, the region in which there are exactly three fixed points does not extend into the desynchronized region. From this we can infer that this region is connected to the desynchronized region by a SNIPER (saddle-node infinite period) bifurcation, and to the other synchronized region by a saddle-node bifurcation. This second region must then be connected to the desynchronized region by a Hopf bifurcation. This structure is qualitatively similar to the bifurcation structure of the forced Kuramoto model with heterogeneous oscillator frequencies \cite{antonsen2008external,childs2008stability}, although the ODEs from which these conditions were derived are quite different. However, the similarity suggests that this type of bifurcation structure could arise for an even broader family of forced coupled oscillator systems. Further work is necessary to understand this connection.

These results also have broader implications to the study of general nonlinear systems. The study of generic models such as amplitude equations, or models derived from center manifold methods, has been a very productive avenue in the study of nonlinear systems due to analytic guarantees of validity. However, care must still be taken, it is important to ensure that the appropriate generic model is applied to the system of interest. In particular, if one is attempting to control a spatial, nonlinear oscillatory medium one must consider the dynamics of the coupling. If the oscillatory medium does not itself diffuse it is possible that the simpler results for controlling the CGLE must be replaced by the more complex rules for controlling the emCGLE, with the possibility of complexities arising due to memory effects in the medium such as amplitude death.

These complications can be of practical importance for control of experimental systems, depending on whether the oscillator medium itself diffuses or the local oscillators are fixed and couple by diffusing an external medium. Examples of the latter case include populations of cells that communicate with oscillating signals or chemical systems such as BZ droplet systems \cite{taylor2009dynamical, toiya2008diffusively, toiya2010synchronization}. In these cases it is possible to gain additional control over the system by controlling the medium. In populations of \emph{Dictyostelium} it is possible to affect the degradation rate of cAMP, the signalling molecule that couples the oscillating cells, in order to change the dynamics of a freely oscillating population (for example by expressing a protein that actively degrades it, or even by washing the medium out) and we show that this mechanism can also be used to affect the susceptibility of such populations to external control.

It is also important to note the importance of boundary conditions in the case of the standard CGLE on the type of synchronization achieved. With realistic, zero-flux boundary conditions several higher order synchronization regimes were observed. This indicates not only that there is a rich bifurcation structure present in the spatially-driven CGLE but that this bifurcation structure is strongly dependent on the boundary conditions of the system. This suggests avenues for future work in analyzing the linear stability of the driven system in these different cases. 

Furthermore, while these models are generically valid for oscillatory systems close to the onset of oscillations, many oscillatory systems in biology are strongly nonlinear in character, for example relaxation oscillations with strong positive feedback. Previous work has shown that phenomena such as amplitude death and memory effects can play an important role in the behavior of such systems \cite{noorbakhsh2015modeling}, but how they will respond to drives is less well understood. 

Our results show that by controlling the spatial variation in a drive, it is possible to vary the frequency range that an oscillatory medium will synchronize to. They also demonstrate that there are universal, qualitative phenomena associated with driving local oscillators coupled by an external medium that these features
may have important practical applications for experimentally controlling complex, biological systems such as cellular populations.

\bibliography{refs}

\begin{thebibliography}{34}%
\makeatletter
\providecommand \@ifxundefined [1]{%
 \@ifx{#1\undefined}
}%
\providecommand \@ifnum [1]{%
 \ifnum #1\expandafter \@firstoftwo
 \else \expandafter \@secondoftwo
 \fi
}%
\providecommand \@ifx [1]{%
 \ifx #1\expandafter \@firstoftwo
 \else \expandafter \@secondoftwo
 \fi
}%
\providecommand \natexlab [1]{#1}%
\providecommand \enquote  [1]{``#1''}%
\providecommand \bibnamefont  [1]{#1}%
\providecommand \bibfnamefont [1]{#1}%
\providecommand \citenamefont [1]{#1}%
\providecommand \href@noop [0]{\@secondoftwo}%
\providecommand \href [0]{\begingroup \@sanitize@url \@href}%
\providecommand \@href[1]{\@@startlink{#1}\@@href}%
\providecommand \@@href[1]{\endgroup#1\@@endlink}%
\providecommand \@sanitize@url [0]{\catcode `\\12\catcode `\$12\catcode
  `\&12\catcode `\#12\catcode `\^12\catcode `\_12\catcode `\%12\relax}%
\providecommand \@@startlink[1]{}%
\providecommand \@@endlink[0]{}%
\providecommand \url  [0]{\begingroup\@sanitize@url \@url }%
\providecommand \@url [1]{\endgroup\@href {#1}{\urlprefix }}%
\providecommand \urlprefix  [0]{URL }%
\providecommand \Eprint [0]{\href }%
\providecommand \doibase [0]{http://dx.doi.org/}%
\providecommand \selectlanguage [0]{\@gobble}%
\providecommand \bibinfo  [0]{\@secondoftwo}%
\providecommand \bibfield  [0]{\@secondoftwo}%
\providecommand \translation [1]{[#1]}%
\providecommand \BibitemOpen [0]{}%
\providecommand \bibitemStop [0]{}%
\providecommand \bibitemNoStop [0]{.\EOS\space}%
\providecommand \EOS [0]{\spacefactor3000\relax}%
\providecommand \BibitemShut  [1]{\csname bibitem#1\endcsname}%
\let\auto@bib@innerbib\@empty
\bibitem [{\citenamefont {Koizumi}(2010)}]{koizumi2010synchronization}%
  \BibitemOpen
  \bibfield  {author} {\bibinfo {author} {\bibfnamefont {S.}~\bibnamefont
  {Koizumi}},\ }\href@noop {} {\bibfield  {journal} {\bibinfo  {journal} {The
  FEBS journal}\ }\textbf {\bibinfo {volume} {277}},\ \bibinfo {pages} {286}
  (\bibinfo {year} {2010})}\BibitemShut {NoStop}%
\bibitem [{\citenamefont {Benninger}\ \emph {et~al.}(2008)\citenamefont
  {Benninger}, \citenamefont {Zhang}, \citenamefont {Head}, \citenamefont
  {Satin},\ and\ \citenamefont {Piston}}]{benninger2008gap}%
  \BibitemOpen
  \bibfield  {author} {\bibinfo {author} {\bibfnamefont {R.~K.}\ \bibnamefont
  {Benninger}}, \bibinfo {author} {\bibfnamefont {M.}~\bibnamefont {Zhang}},
  \bibinfo {author} {\bibfnamefont {W.~S.}\ \bibnamefont {Head}}, \bibinfo
  {author} {\bibfnamefont {L.~S.}\ \bibnamefont {Satin}}, \ and\ \bibinfo
  {author} {\bibfnamefont {D.~W.}\ \bibnamefont {Piston}},\ }\href@noop {}
  {\bibfield  {journal} {\bibinfo  {journal} {Biophysical journal}\ }\textbf
  {\bibinfo {volume} {95}},\ \bibinfo {pages} {5048} (\bibinfo {year}
  {2008})}\BibitemShut {NoStop}%
\bibitem [{\citenamefont {Ito}\ \emph {et~al.}(2007)\citenamefont {Ito},
  \citenamefont {Kageyama}, \citenamefont {Mutsuda}, \citenamefont {Nakajima},
  \citenamefont {Oyama},\ and\ \citenamefont {Kondo}}]{ito2007autonomous}%
  \BibitemOpen
  \bibfield  {author} {\bibinfo {author} {\bibfnamefont {H.}~\bibnamefont
  {Ito}}, \bibinfo {author} {\bibfnamefont {H.}~\bibnamefont {Kageyama}},
  \bibinfo {author} {\bibfnamefont {M.}~\bibnamefont {Mutsuda}}, \bibinfo
  {author} {\bibfnamefont {M.}~\bibnamefont {Nakajima}}, \bibinfo {author}
  {\bibfnamefont {T.}~\bibnamefont {Oyama}}, \ and\ \bibinfo {author}
  {\bibfnamefont {T.}~\bibnamefont {Kondo}},\ }\href@noop {} {\bibfield
  {journal} {\bibinfo  {journal} {Nature structural \& molecular biology}\
  }\textbf {\bibinfo {volume} {14}},\ \bibinfo {pages} {1084} (\bibinfo {year}
  {2007})}\BibitemShut {NoStop}%
\bibitem [{\citenamefont {van Zon}\ \emph {et~al.}(2007)\citenamefont {van
  Zon}, \citenamefont {Lubensky}, \citenamefont {Altena},\ and\ \citenamefont
  {ten Wolde}}]{van2007allosteric}%
  \BibitemOpen
  \bibfield  {author} {\bibinfo {author} {\bibfnamefont {J.~S.}\ \bibnamefont
  {van Zon}}, \bibinfo {author} {\bibfnamefont {D.~K.}\ \bibnamefont
  {Lubensky}}, \bibinfo {author} {\bibfnamefont {P.~R.}\ \bibnamefont
  {Altena}}, \ and\ \bibinfo {author} {\bibfnamefont {P.~R.}\ \bibnamefont {ten
  Wolde}},\ }\href@noop {} {\bibfield  {journal} {\bibinfo  {journal}
  {Proceedings of the National Academy of Sciences}\ }\textbf {\bibinfo
  {volume} {104}},\ \bibinfo {pages} {7420} (\bibinfo {year}
  {2007})}\BibitemShut {NoStop}%
\bibitem [{\citenamefont {Dan{\o}}\ \emph {et~al.}(1999)\citenamefont
  {Dan{\o}}, \citenamefont {S{\o}rensen},\ and\ \citenamefont
  {Hynne}}]{dano1999sustained}%
  \BibitemOpen
  \bibfield  {author} {\bibinfo {author} {\bibfnamefont {S.}~\bibnamefont
  {Dan{\o}}}, \bibinfo {author} {\bibfnamefont {P.~G.}\ \bibnamefont
  {S{\o}rensen}}, \ and\ \bibinfo {author} {\bibfnamefont {F.}~\bibnamefont
  {Hynne}},\ }\href@noop {} {\bibfield  {journal} {\bibinfo  {journal}
  {Nature}\ }\textbf {\bibinfo {volume} {402}},\ \bibinfo {pages} {320}
  (\bibinfo {year} {1999})}\BibitemShut {NoStop}%
\bibitem [{\citenamefont {Gregor}\ \emph {et~al.}(2010)\citenamefont {Gregor},
  \citenamefont {Fujimoto}, \citenamefont {Masaki},\ and\ \citenamefont
  {Sawai}}]{gregor2010onset}%
  \BibitemOpen
  \bibfield  {author} {\bibinfo {author} {\bibfnamefont {T.}~\bibnamefont
  {Gregor}}, \bibinfo {author} {\bibfnamefont {K.}~\bibnamefont {Fujimoto}},
  \bibinfo {author} {\bibfnamefont {N.}~\bibnamefont {Masaki}}, \ and\ \bibinfo
  {author} {\bibfnamefont {S.}~\bibnamefont {Sawai}},\ }\href@noop {}
  {\bibfield  {journal} {\bibinfo  {journal} {Science}\ }\textbf {\bibinfo
  {volume} {328}},\ \bibinfo {pages} {1021} (\bibinfo {year}
  {2010})}\BibitemShut {NoStop}%
\bibitem [{\citenamefont {De~Monte}\ \emph {et~al.}(2007)\citenamefont
  {De~Monte}, \citenamefont {d'Ovidio}, \citenamefont {Dan{\o}},\ and\
  \citenamefont {S{\o}rensen}}]{de2007dynamical}%
  \BibitemOpen
  \bibfield  {author} {\bibinfo {author} {\bibfnamefont {S.}~\bibnamefont
  {De~Monte}}, \bibinfo {author} {\bibfnamefont {F.}~\bibnamefont {d'Ovidio}},
  \bibinfo {author} {\bibfnamefont {S.}~\bibnamefont {Dan{\o}}}, \ and\
  \bibinfo {author} {\bibfnamefont {P.~G.}\ \bibnamefont {S{\o}rensen}},\
  }\href@noop {} {\bibfield  {journal} {\bibinfo  {journal} {Proceedings of the
  National Academy of Sciences}\ }\textbf {\bibinfo {volume} {104}},\ \bibinfo
  {pages} {18377} (\bibinfo {year} {2007})}\BibitemShut {NoStop}%
\bibitem [{\citenamefont {Huygens}\ and\ \citenamefont
  {Muguet}()}]{huygens1673christiani}%
  \BibitemOpen
  \bibfield  {author} {\bibinfo {author} {\bibfnamefont {C.}~\bibnamefont
  {Huygens}}\ and\ \bibinfo {author} {\bibfnamefont {F.}~\bibnamefont
  {Muguet}},\ }\href@noop {} {\bibfield  {journal} {\bibinfo  {journal} {F.
  Horologium oscillatorium sive De motu pendulorum ad horologia aptato
  demonstrationes geometricae. Parisiis: Apud F. Muguet, Paris, France}\
  }\textbf {\bibinfo {volume} {1673}}}\BibitemShut {NoStop}%
\bibitem [{\citenamefont {Aranson}\ and\ \citenamefont
  {Kramer}(2002)}]{aranson2002world}%
  \BibitemOpen
  \bibfield  {author} {\bibinfo {author} {\bibfnamefont {I.~S.}\ \bibnamefont
  {Aranson}}\ and\ \bibinfo {author} {\bibfnamefont {L.}~\bibnamefont
  {Kramer}},\ }\href@noop {} {\bibfield  {journal} {\bibinfo  {journal}
  {Reviews of modern physics}\ }\textbf {\bibinfo {volume} {74}},\ \bibinfo
  {pages} {99} (\bibinfo {year} {2002})}\BibitemShut {NoStop}%
\bibitem [{\citenamefont {Liu}\ \emph {et~al.}(2015)\citenamefont {Liu},
  \citenamefont {Prindle}, \citenamefont {Humphries}, \citenamefont
  {Gabalda-Sagarra}, \citenamefont {Asally}, \citenamefont {Dong-yeon},
  \citenamefont {Ly}, \citenamefont {Garcia-Ojalvo},\ and\ \citenamefont
  {S{\"u}el}}]{liu2015metabolic}%
  \BibitemOpen
  \bibfield  {author} {\bibinfo {author} {\bibfnamefont {J.}~\bibnamefont
  {Liu}}, \bibinfo {author} {\bibfnamefont {A.}~\bibnamefont {Prindle}},
  \bibinfo {author} {\bibfnamefont {J.}~\bibnamefont {Humphries}}, \bibinfo
  {author} {\bibfnamefont {M.}~\bibnamefont {Gabalda-Sagarra}}, \bibinfo
  {author} {\bibfnamefont {M.}~\bibnamefont {Asally}}, \bibinfo {author}
  {\bibfnamefont {D.~L.}\ \bibnamefont {Dong-yeon}}, \bibinfo {author}
  {\bibfnamefont {S.}~\bibnamefont {Ly}}, \bibinfo {author} {\bibfnamefont
  {J.}~\bibnamefont {Garcia-Ojalvo}}, \ and\ \bibinfo {author} {\bibfnamefont
  {G.~M.}\ \bibnamefont {S{\"u}el}},\ }\href@noop {} {\bibfield  {journal}
  {\bibinfo  {journal} {Nature}\ }\textbf {\bibinfo {volume} {523}},\ \bibinfo
  {pages} {550} (\bibinfo {year} {2015})}\BibitemShut {NoStop}%
\bibitem [{\citenamefont {Liu}\ \emph {et~al.}(2017)\citenamefont {Liu},
  \citenamefont {Martinez-Corral}, \citenamefont {Prindle}, \citenamefont
  {Dong-yeon}, \citenamefont {Larkin}, \citenamefont {Gabalda-Sagarra},
  \citenamefont {Garcia-Ojalvo},\ and\ \citenamefont
  {S{\"u}el}}]{liu2017coupling}%
  \BibitemOpen
  \bibfield  {author} {\bibinfo {author} {\bibfnamefont {J.}~\bibnamefont
  {Liu}}, \bibinfo {author} {\bibfnamefont {R.}~\bibnamefont
  {Martinez-Corral}}, \bibinfo {author} {\bibfnamefont {A.}~\bibnamefont
  {Prindle}}, \bibinfo {author} {\bibfnamefont {D.~L.}\ \bibnamefont
  {Dong-yeon}}, \bibinfo {author} {\bibfnamefont {J.}~\bibnamefont {Larkin}},
  \bibinfo {author} {\bibfnamefont {M.}~\bibnamefont {Gabalda-Sagarra}},
  \bibinfo {author} {\bibfnamefont {J.}~\bibnamefont {Garcia-Ojalvo}}, \ and\
  \bibinfo {author} {\bibfnamefont {G.~M.}\ \bibnamefont {S{\"u}el}},\
  }\href@noop {} {\bibfield  {journal} {\bibinfo  {journal} {Science}\ }\textbf
  {\bibinfo {volume} {356}},\ \bibinfo {pages} {638} (\bibinfo {year}
  {2017})}\BibitemShut {NoStop}%
\bibitem [{\citenamefont {Cross}\ and\ \citenamefont
  {Hohenberg}(1993)}]{cross1993pattern}%
  \BibitemOpen
  \bibfield  {author} {\bibinfo {author} {\bibfnamefont {M.~C.}\ \bibnamefont
  {Cross}}\ and\ \bibinfo {author} {\bibfnamefont {P.~C.}\ \bibnamefont
  {Hohenberg}},\ }\href@noop {} {\bibfield  {journal} {\bibinfo  {journal}
  {Reviews of modern physics}\ }\textbf {\bibinfo {volume} {65}},\ \bibinfo
  {pages} {851} (\bibinfo {year} {1993})}\BibitemShut {NoStop}%
\bibitem [{\citenamefont {Strogatz}(1998)}]{strogatz1998death}%
  \BibitemOpen
  \bibfield  {author} {\bibinfo {author} {\bibfnamefont {S.~H.}\ \bibnamefont
  {Strogatz}},\ }\href@noop {} {\bibfield  {journal} {\bibinfo  {journal}
  {Nature}\ }\textbf {\bibinfo {volume} {394}},\ \bibinfo {pages} {316}
  (\bibinfo {year} {1998})}\BibitemShut {NoStop}%
\bibitem [{\citenamefont {Schwab}\ \emph
  {et~al.}(2012{\natexlab{a}})\citenamefont {Schwab}, \citenamefont {Baetica},\
  and\ \citenamefont {Mehta}}]{schwab2012dynamical}%
  \BibitemOpen
  \bibfield  {author} {\bibinfo {author} {\bibfnamefont {D.~J.}\ \bibnamefont
  {Schwab}}, \bibinfo {author} {\bibfnamefont {A.}~\bibnamefont {Baetica}}, \
  and\ \bibinfo {author} {\bibfnamefont {P.}~\bibnamefont {Mehta}},\
  }\href@noop {} {\bibfield  {journal} {\bibinfo  {journal} {Physica D:
  Nonlinear Phenomena}\ }\textbf {\bibinfo {volume} {241}},\ \bibinfo {pages}
  {1782} (\bibinfo {year} {2012}{\natexlab{a}})}\BibitemShut {NoStop}%
\bibitem [{\citenamefont {Schwab}\ \emph
  {et~al.}(2012{\natexlab{b}})\citenamefont {Schwab}, \citenamefont {Plunk},\
  and\ \citenamefont {Mehta}}]{schwab2012kuramoto}%
  \BibitemOpen
  \bibfield  {author} {\bibinfo {author} {\bibfnamefont {D.~J.}\ \bibnamefont
  {Schwab}}, \bibinfo {author} {\bibfnamefont {G.~G.}\ \bibnamefont {Plunk}}, \
  and\ \bibinfo {author} {\bibfnamefont {P.}~\bibnamefont {Mehta}},\
  }\href@noop {} {\bibfield  {journal} {\bibinfo  {journal} {Chaos: An
  Interdisciplinary Journal of Nonlinear Science}\ }\textbf {\bibinfo {volume}
  {22}},\ \bibinfo {pages} {043139} (\bibinfo {year}
  {2012}{\natexlab{b}})}\BibitemShut {NoStop}%
\bibitem [{\citenamefont {Noorbakhsh}\ \emph {et~al.}(2015)\citenamefont
  {Noorbakhsh}, \citenamefont {Schwab}, \citenamefont {Sgro}, \citenamefont
  {Gregor},\ and\ \citenamefont {Mehta}}]{noorbakhsh2015modeling}%
  \BibitemOpen
  \bibfield  {author} {\bibinfo {author} {\bibfnamefont {J.}~\bibnamefont
  {Noorbakhsh}}, \bibinfo {author} {\bibfnamefont {D.~J.}\ \bibnamefont
  {Schwab}}, \bibinfo {author} {\bibfnamefont {A.~E.}\ \bibnamefont {Sgro}},
  \bibinfo {author} {\bibfnamefont {T.}~\bibnamefont {Gregor}}, \ and\ \bibinfo
  {author} {\bibfnamefont {P.}~\bibnamefont {Mehta}},\ }\href@noop {}
  {\bibfield  {journal} {\bibinfo  {journal} {Physical Review E}\ }\textbf
  {\bibinfo {volume} {91}},\ \bibinfo {pages} {062711} (\bibinfo {year}
  {2015})}\BibitemShut {NoStop}%
\bibitem [{\citenamefont {Tanaka}\ and\ \citenamefont
  {Kuramoto}(2003)}]{tanaka2003complex}%
  \BibitemOpen
  \bibfield  {author} {\bibinfo {author} {\bibfnamefont {D.}~\bibnamefont
  {Tanaka}}\ and\ \bibinfo {author} {\bibfnamefont {Y.}~\bibnamefont
  {Kuramoto}},\ }\href@noop {} {\bibfield  {journal} {\bibinfo  {journal}
  {Physical Review E}\ }\textbf {\bibinfo {volume} {68}},\ \bibinfo {pages}
  {026219} (\bibinfo {year} {2003})}\BibitemShut {NoStop}%
\bibitem [{\citenamefont {Kuramoto}(1997)}]{kuramoto1997phase}%
  \BibitemOpen
  \bibfield  {author} {\bibinfo {author} {\bibfnamefont {Y.}~\bibnamefont
  {Kuramoto}},\ }\href@noop {} {\bibfield  {journal} {\bibinfo  {journal}
  {International Journal of Bifurcation and Chaos}\ }\textbf {\bibinfo {volume}
  {7}},\ \bibinfo {pages} {789} (\bibinfo {year} {1997})}\BibitemShut {NoStop}%
\bibitem [{\citenamefont {Chat{\'e}}\ \emph {et~al.}(1999)\citenamefont
  {Chat{\'e}}, \citenamefont {Pikovsky},\ and\ \citenamefont
  {Rudzick}}]{chate1999forcing}%
  \BibitemOpen
  \bibfield  {author} {\bibinfo {author} {\bibfnamefont {H.}~\bibnamefont
  {Chat{\'e}}}, \bibinfo {author} {\bibfnamefont {A.}~\bibnamefont {Pikovsky}},
  \ and\ \bibinfo {author} {\bibfnamefont {O.}~\bibnamefont {Rudzick}},\
  }\href@noop {} {\bibfield  {journal} {\bibinfo  {journal} {Physica D:
  Nonlinear Phenomena}\ }\textbf {\bibinfo {volume} {131}},\ \bibinfo {pages}
  {17} (\bibinfo {year} {1999})}\BibitemShut {NoStop}%
\bibitem [{\citenamefont {Newell}\ and\ \citenamefont
  {AC}(1974)}]{newell1974envelope}%
  \BibitemOpen
  \bibfield  {author} {\bibinfo {author} {\bibfnamefont {A.~C.}\ \bibnamefont
  {Newell}}\ and\ \bibinfo {author} {\bibfnamefont {N.}~\bibnamefont {AC}},\
  }\href@noop {} {\  (\bibinfo {year} {1974})}\BibitemShut {NoStop}%
\bibitem [{\citenamefont {Kuramoto}(2003)}]{kuramoto2003chemical}%
  \BibitemOpen
  \bibfield  {author} {\bibinfo {author} {\bibfnamefont {Y.}~\bibnamefont
  {Kuramoto}},\ }\href@noop {} {\emph {\bibinfo {title} {Chemical oscillations,
  waves, and turbulence}}}\ (\bibinfo  {publisher} {Courier Corporation},\
  \bibinfo {year} {2003})\BibitemShut {NoStop}%
\bibitem [{\citenamefont {van Hecke~M}\ \emph {et~al.}(1994)\citenamefont {van
  Hecke~M}, \citenamefont {PC},\ and\ \citenamefont {van
  Saarloos~W}}]{vanHecke1994Amplitude}%
  \BibitemOpen
  \bibfield  {author} {\bibinfo {author} {\bibnamefont {van Hecke~M}}, \bibinfo
  {author} {\bibfnamefont {H.}~\bibnamefont {PC}}, \ and\ \bibinfo {author}
  {\bibnamefont {van Saarloos~W}},\ }in\ \href@noop {} {\emph {\bibinfo
  {booktitle} {Fundamental Problems in Statistical Mechanics VIII}}},\ \bibinfo
  {editor} {edited by\ \bibinfo {editor} {\bibfnamefont {H.}~\bibnamefont {van
  Beijeren}}\ and\ \bibinfo {editor} {\bibfnamefont {M.}~\bibnamefont
  {Ernst}}}\ (\bibinfo  {publisher} {Elsevier},\ \bibinfo {year}
  {1994})\BibitemShut {NoStop}%
\bibitem [{\citenamefont {Pismen}\ \emph {et~al.}(1999)\citenamefont {Pismen},
  \citenamefont {Pismen} \emph {et~al.}}]{pismen1999vortices}%
  \BibitemOpen
  \bibfield  {author} {\bibinfo {author} {\bibfnamefont {L.~M.}\ \bibnamefont
  {Pismen}}, \bibinfo {author} {\bibfnamefont {L.~M.}\ \bibnamefont {Pismen}},
  \emph {et~al.},\ }\href@noop {} {\emph {\bibinfo {title} {Vortices in
  nonlinear fields: From liquid crystals to superfluids, from non-equilibrium
  patterns to cosmic strings}}},\ Vol.\ \bibinfo {volume} {100}\ (\bibinfo
  {publisher} {Oxford University Press},\ \bibinfo {year} {1999})\BibitemShut
  {NoStop}%
\bibitem [{\citenamefont {Ouyang}\ and\ \citenamefont
  {Flesselles}(1996)}]{ouyang1996transition}%
  \BibitemOpen
  \bibfield  {author} {\bibinfo {author} {\bibfnamefont {Q.}~\bibnamefont
  {Ouyang}}\ and\ \bibinfo {author} {\bibfnamefont {J.-M.}\ \bibnamefont
  {Flesselles}},\ }\href@noop {} {\bibfield  {journal} {\bibinfo  {journal}
  {Nature}\ }\textbf {\bibinfo {volume} {379}},\ \bibinfo {pages} {143}
  (\bibinfo {year} {1996})}\BibitemShut {NoStop}%
\bibitem [{\citenamefont {Leweke}\ and\ \citenamefont
  {Provansal}(1994)}]{leweke1994model}%
  \BibitemOpen
  \bibfield  {author} {\bibinfo {author} {\bibfnamefont {T.}~\bibnamefont
  {Leweke}}\ and\ \bibinfo {author} {\bibfnamefont {M.}~\bibnamefont
  {Provansal}},\ }\href@noop {} {\bibfield  {journal} {\bibinfo  {journal}
  {Physical review letters}\ }\textbf {\bibinfo {volume} {72}},\ \bibinfo
  {pages} {3174} (\bibinfo {year} {1994})}\BibitemShut {NoStop}%
\bibitem [{\citenamefont {Garc{\'\i}a-Morales}\ and\ \citenamefont
  {Krischer}(2012)}]{garcia2012complex}%
  \BibitemOpen
  \bibfield  {author} {\bibinfo {author} {\bibfnamefont {V.}~\bibnamefont
  {Garc{\'\i}a-Morales}}\ and\ \bibinfo {author} {\bibfnamefont
  {K.}~\bibnamefont {Krischer}},\ }\href@noop {} {\bibfield  {journal}
  {\bibinfo  {journal} {Contemporary Physics}\ }\textbf {\bibinfo {volume}
  {53}},\ \bibinfo {pages} {79} (\bibinfo {year} {2012})}\BibitemShut {NoStop}%
\bibitem [{\citenamefont {Antonsen~Jr}\ \emph {et~al.}(2008)\citenamefont
  {Antonsen~Jr}, \citenamefont {Faghih}, \citenamefont {Girvan}, \citenamefont
  {Ott},\ and\ \citenamefont {Platig}}]{antonsen2008external}%
  \BibitemOpen
  \bibfield  {author} {\bibinfo {author} {\bibfnamefont {T.}~\bibnamefont
  {Antonsen~Jr}}, \bibinfo {author} {\bibfnamefont {R.}~\bibnamefont {Faghih}},
  \bibinfo {author} {\bibfnamefont {M.}~\bibnamefont {Girvan}}, \bibinfo
  {author} {\bibfnamefont {E.}~\bibnamefont {Ott}}, \ and\ \bibinfo {author}
  {\bibfnamefont {J.}~\bibnamefont {Platig}},\ }\href@noop {} {\bibfield
  {journal} {\bibinfo  {journal} {Chaos: An Interdisciplinary Journal of
  Nonlinear Science}\ }\textbf {\bibinfo {volume} {18}},\ \bibinfo {pages}
  {037112} (\bibinfo {year} {2008})}\BibitemShut {NoStop}%
\bibitem [{\citenamefont {Childs}\ and\ \citenamefont
  {Strogatz}(2008)}]{childs2008stability}%
  \BibitemOpen
  \bibfield  {author} {\bibinfo {author} {\bibfnamefont {L.~M.}\ \bibnamefont
  {Childs}}\ and\ \bibinfo {author} {\bibfnamefont {S.~H.}\ \bibnamefont
  {Strogatz}},\ }\href@noop {} {\bibfield  {journal} {\bibinfo  {journal}
  {Chaos: An Interdisciplinary Journal of Nonlinear Science}\ }\textbf
  {\bibinfo {volume} {18}},\ \bibinfo {pages} {043128} (\bibinfo {year}
  {2008})}\BibitemShut {NoStop}%
\bibitem [{\citenamefont {Garc{\'\i}a-{\'A}lvarez}\ \emph
  {et~al.}(2008)\citenamefont {Garc{\'\i}a-{\'A}lvarez}, \citenamefont
  {Stefanovska},\ and\ \citenamefont {McClintock}}]{garcia2008high}%
  \BibitemOpen
  \bibfield  {author} {\bibinfo {author} {\bibfnamefont {D.}~\bibnamefont
  {Garc{\'\i}a-{\'A}lvarez}}, \bibinfo {author} {\bibfnamefont
  {A.}~\bibnamefont {Stefanovska}}, \ and\ \bibinfo {author} {\bibfnamefont
  {P.~V.}\ \bibnamefont {McClintock}},\ }\href@noop {} {\bibfield  {journal}
  {\bibinfo  {journal} {Physical Review E}\ }\textbf {\bibinfo {volume} {77}},\
  \bibinfo {pages} {056203} (\bibinfo {year} {2008})}\BibitemShut {NoStop}%
\bibitem [{\citenamefont {McMains}\ \emph {et~al.}(2008)\citenamefont
  {McMains}, \citenamefont {Liao},\ and\ \citenamefont
  {Kimmel}}]{mcmains2008oscillatory}%
  \BibitemOpen
  \bibfield  {author} {\bibinfo {author} {\bibfnamefont {V.~C.}\ \bibnamefont
  {McMains}}, \bibinfo {author} {\bibfnamefont {X.-H.}\ \bibnamefont {Liao}}, \
  and\ \bibinfo {author} {\bibfnamefont {A.~R.}\ \bibnamefont {Kimmel}},\
  }\href@noop {} {\bibfield  {journal} {\bibinfo  {journal} {Ageing research
  reviews}\ }\textbf {\bibinfo {volume} {7}},\ \bibinfo {pages} {234} (\bibinfo
  {year} {2008})}\BibitemShut {NoStop}%
\bibitem [{\citenamefont {Sgro}\ \emph {et~al.}(2015)\citenamefont {Sgro},
  \citenamefont {Schwab}, \citenamefont {Noorbakhsh}, \citenamefont {Mestler},
  \citenamefont {Mehta},\ and\ \citenamefont {Gregor}}]{sgro2015intracellular}%
  \BibitemOpen
  \bibfield  {author} {\bibinfo {author} {\bibfnamefont {A.~E.}\ \bibnamefont
  {Sgro}}, \bibinfo {author} {\bibfnamefont {D.~J.}\ \bibnamefont {Schwab}},
  \bibinfo {author} {\bibfnamefont {J.}~\bibnamefont {Noorbakhsh}}, \bibinfo
  {author} {\bibfnamefont {T.}~\bibnamefont {Mestler}}, \bibinfo {author}
  {\bibfnamefont {P.}~\bibnamefont {Mehta}}, \ and\ \bibinfo {author}
  {\bibfnamefont {T.}~\bibnamefont {Gregor}},\ }\href@noop {} {\bibfield
  {journal} {\bibinfo  {journal} {Molecular systems biology}\ }\textbf
  {\bibinfo {volume} {11}},\ \bibinfo {pages} {779} (\bibinfo {year}
  {2015})}\BibitemShut {NoStop}%
\bibitem [{\citenamefont {Taylor}\ \emph {et~al.}(2009)\citenamefont {Taylor},
  \citenamefont {Tinsley}, \citenamefont {Wang}, \citenamefont {Huang},\ and\
  \citenamefont {Showalter}}]{taylor2009dynamical}%
  \BibitemOpen
  \bibfield  {author} {\bibinfo {author} {\bibfnamefont {A.~F.}\ \bibnamefont
  {Taylor}}, \bibinfo {author} {\bibfnamefont {M.~R.}\ \bibnamefont {Tinsley}},
  \bibinfo {author} {\bibfnamefont {F.}~\bibnamefont {Wang}}, \bibinfo {author}
  {\bibfnamefont {Z.}~\bibnamefont {Huang}}, \ and\ \bibinfo {author}
  {\bibfnamefont {K.}~\bibnamefont {Showalter}},\ }\href@noop {} {\bibfield
  {journal} {\bibinfo  {journal} {Science}\ }\textbf {\bibinfo {volume}
  {323}},\ \bibinfo {pages} {614} (\bibinfo {year} {2009})}\BibitemShut
  {NoStop}%
\bibitem [{\citenamefont {Toiya}\ \emph {et~al.}(2008)\citenamefont {Toiya},
  \citenamefont {Vanag},\ and\ \citenamefont {Epstein}}]{toiya2008diffusively}%
  \BibitemOpen
  \bibfield  {author} {\bibinfo {author} {\bibfnamefont {M.}~\bibnamefont
  {Toiya}}, \bibinfo {author} {\bibfnamefont {V.~K.}\ \bibnamefont {Vanag}}, \
  and\ \bibinfo {author} {\bibfnamefont {I.~R.}\ \bibnamefont {Epstein}},\
  }\href@noop {} {\bibfield  {journal} {\bibinfo  {journal} {Angewandte
  Chemie}\ }\textbf {\bibinfo {volume} {120}},\ \bibinfo {pages} {7867}
  (\bibinfo {year} {2008})}\BibitemShut {NoStop}%
\bibitem [{\citenamefont {Toiya}\ \emph {et~al.}(2010)\citenamefont {Toiya},
  \citenamefont {Gonz{\'a}lez-Ochoa}, \citenamefont {Vanag}, \citenamefont
  {Fraden},\ and\ \citenamefont {Epstein}}]{toiya2010synchronization}%
  \BibitemOpen
  \bibfield  {author} {\bibinfo {author} {\bibfnamefont {M.}~\bibnamefont
  {Toiya}}, \bibinfo {author} {\bibfnamefont {H.~O.}\ \bibnamefont
  {Gonz{\'a}lez-Ochoa}}, \bibinfo {author} {\bibfnamefont {V.~K.}\ \bibnamefont
  {Vanag}}, \bibinfo {author} {\bibfnamefont {S.}~\bibnamefont {Fraden}}, \
  and\ \bibinfo {author} {\bibfnamefont {I.~R.}\ \bibnamefont {Epstein}},\
  }\href@noop {} {\bibfield  {journal} {\bibinfo  {journal} {The Journal of
  Physical Chemistry Letters}\ }\textbf {\bibinfo {volume} {1}},\ \bibinfo
  {pages} {1241} (\bibinfo {year} {2010})}\BibitemShut {NoStop}%
\end{thebibliography}%

\end{document}